\documentclass[conference]{IEEEtran}
\IEEEoverridecommandlockouts
\usepackage{cite}
\usepackage{amsmath,amssymb,amsfonts}
\usepackage{algorithmic}
\usepackage{graphicx}
\usepackage{textcomp}
\usepackage{xcolor}
\usepackage{todonotes}
\usepackage{framed}
\usepackage{float}
\usepackage[export]{adjustbox}
\usepackage{caption}
\usepackage{subcaption}

\usepackage[colorlinks=true, linkcolor=black,urlcolor=black, citecolor=black]{hyperref}
\def\BibTeX{{\rm B\kern-.05em{\sc i\kern-.025em b}\kern-.08em
    T\kern-.1667em\lower.7ex\hbox{E}\kern-.125emX}}
\begin{document}

\title{Determining the Difficulties of Students With Dyslexia via Virtual Reality and Artificial Intelligence: An Exploratory Analysis \\
}

\newcommand{\newlineauthors}{%
  \end{@IEEEauthorhalign}\hfill\mbox{}\par
  \mbox{}\hfill\begin{@IEEEauthorhalign}
}
\author{\IEEEauthorblockN{Enrique Yeguas-Bolívar}
\IEEEauthorblockA{\textit {Computing and Numerical Analysis} \\
\textit{University of Córdoba}\\
Córdoba, Spain \\
eyeguas@uco.es}
\and
\IEEEauthorblockN{José M. Alcalde-Llergo}
\IEEEauthorblockA{\textit{Artificial Vision Applications (AVA)} \\
\textit{University of Córdoba}\\
Córdoba, Spain \\
i72alllj@uco.es}
\and
\IEEEauthorblockN{Pilar Aparicio-Martínez}
\IEEEauthorblockA{\textit{Nursing, Physiotherapy and Pharmacology} \\
\textit{University of Córdoba}\\
Córdoba, Spain \\
n32apmap@uco.es}
\and
\IEEEauthorblockN{Juri Taborri}
\IEEEauthorblockA{\textit{Dept. of Economics, Engineering,} \\
\textit{Society and Business Organization (DEIM)}\\
\textit{University of Tuscia}\\
Viterbo, Italy \\
juri.taborri@unitus.it}
\and
\IEEEauthorblockN{Andrea Zingoni}
\IEEEauthorblockA{\textit{Dept. of Economics, Engineering,} \\
\textit{Society and Business Organization (DEIM)}\\
\textit{University of Tuscia}\\
Viterbo, Italy \\
andrea.zingoni@unitus.it}

\and
\IEEEauthorblockN{Sara Pinzi}
\IEEEauthorblockA{\textit {Dept. Physical Chemistry} \\
\textit{and Applied Thermodynamics}\\
\textit{University of Córdoba}\\
Córdoba, Spain \\
qf1pinps@uco.es}
}

\maketitle

\begin{abstract}
Learning disorders are neurological conditions that affect the brain’s ability to interconnect communication areas. Dyslexic students experience problems with reading, memorizing, and exposing concepts; however the magnitude of these can be mitigated through both therapies and the creation of compensatory mechanisms. Several efforts have been made to mitigate these issues, leading to the creation of digital resources for students with specific learning disorders attending primary and secondary education levels. Conversely, a standard approach is still missed in higher education. The VRAIlexia project has been created to tackle this issue by proposing two different tools: a mobile application integrating virtual reality (VR) to collect data quickly and easily, and an artificial intelligence-based software (AI) to analyze the collected data for customizing the supporting methodology for each student. The first one has been created and is being distributed among dyslexic students in Higher Education Institutions, for the conduction of specific psychological and psychometric tests. The second tool applies specific artificial intelligence algorithms to the data gathered via the application and other surveys. These AI techniques have allowed us to identify the most relevant difficulties faced by the students’ cohort. Our different models have obtained around 90\% mean accuracy for predicting the support tools and learning strategies.
\end{abstract}


\section{Introduction}
Learning disorders are neurological conditions that affect the brain’s ability to interconnect communication areas, worsening the information received, processed, and even the way it is sent. In this heterogeneous group, dyslexia is the most prevalent neurological-based disorder that influences the communication between people. In fact, within these types of disorders, approximately 80\% of cases are reading difficulties~\cite{1}. It is estimated that one out of ten children has reading disorders, increasing the risk of developing emotional or mental issues during childhood, adolescence, and adulthood \cite{2}. In addition, a recent published study aiming at clustering dyslexic profiles in Italian universities showed as an early diagnosis and the use of compensatory tools allow to improve the scholastic and academic career of students \cite{juri}. 

Dyslexic students experience problems with reading, memorizing, and exposing concepts, but the magnitude of these can be mitigated through therapies and the creation of compensatory mechanisms. In recent years, the inclusion of technologies has been growing, especially the ones related to virtual environments, as therapeutical resources \cite{2}. However, its full potential has not yet been used systematically in the above-mentioned field, especially when we refer to the use of techniques such as virtual reality (VR) and artificial intelligence (AI) \cite{3,4}.

Several efforts have been made to mitigate these issues, leading to the creation of digital resources for students with specific learning disorders attending primary and secondary education levels. Conversely, a standard approach is still missed in higher education \cite{2,3}. One example of the inclusion of VR and technology for dyslexic students is the FORDYSVAR project~\cite{fordysvar}, which was focused on designing and creating a virtual world with diverse layers for helping these students in their learning process as well as developing further compensatory mechanisms \cite{1}.

Despite this recent project, no other investigation has focused on higher education institutions (HEIs) and the possible use of AI for determining the specific difficulties. This is concerning since students with dyslexia in HEIs have a low rate of graduating (2\%) and may experience multiple issues~\cite{4}. 

Therefore, the current paper presents the combination and structure of “VRAIlexia”, a project that aims at exploiting VR and AI jointly, to provide support to dyslexic students during their academic career. Section II presents the objectives of our study, Section III introduces part of the VR results consisting in a VR mobile app to conduct psychometric tests for dyslexic students, Section IV shows and analyses the preliminary AI results: datasets, models and prediction of the most suitable support tools and learning strategies for the students and, finally, Section V draws the Conclusions and Future work.

\section{Framework and objectives}
Currently, some quite promising tests have already been carried out with multiple automatic learning techniques, or machine learning \cite{5}, from which the first models have already been trained and prepared for use. Even so, the inclusion of the data from the virtual reality application within the training of the models is still pending \cite{6,7}. Besides, previous work has indicated the creation of models based on testing diverse tools and strategies \cite{8}. Therefore, the current presented research has an objective to obtain different models capable of deciding which study tools and methodologies would be the most effective for each specific student with dyslexia and to identify the mechanisms of compensation developed by higher education level students experiencing this disorder. To do this, the model is trained from a data set made up of the results obtained by the student in a VR application while performing psychological and psychometric tests, combined with subjective information from surveys \cite{1} carried out on different European university students with dyslexia. 

Based on such objectives, the results of the current paper are divided into two different products; first the VR app, based on a mobile application, which includes the sections and tests embedded in the called “Out of the Box”. Secondly, the different models constructed with well-known algorithms, which have been previously used for the analysis of the data obtained via the application, and the most adequate generated models which have been used to predict the objective classes: best tools and learning strategies for dyslexic students.

\section{VRAILEXIA software}

In this section, the description of the VR application to collect data and the AI-based software to assist dyslexic students are presented.

\subsection{VR app: Out of the box}

There are significant and empirical evidences of the beneficial use of virtual environments~\cite{virtualenvironments,virtualenvironments2,fordysvar,virtualenvironments4} for the dyslexic community. Assessment, screening, individual intervention, awareness and empathy
can take place in multisensory, controlled and dynamic virtual environments that provide successful performance and interaction for dyslexics.

Immersion, presence, interaction, transduction and conceptual change are the key characteristics of VR. These characteristics are reported
to contribute to the effectiveness in order to help parents and teachers with dyslexic teenagers and students understand the cognitive dyslexia challenges~\cite{virtualenvironments3}. 
On the other hand, using VR improves the quality of data gathered from dyslexic students and saves time as concerns data collected by the usual mechanisms~\cite{1}. The use of virtual environments seeks to increase engagement, accessibility and concentration in the interaction whilst building on the strengths of the current devices.

The VR experiences can be very abstract, and often difficult to visualize, let alone clarify to another person. Prototyping development is useful in VR since a simple prototype can make sense of the idea for the different psychometric tests. Prototyping can be one of the most valuable tools that a VR designer can materialize throughout their design process. 

We really need cases for dyslexic students where VR makes sense and provides value: difficulty to concentrate when studying, distractions during reading, absence of a dedicated workspace, etc. It is difficult for many dyslexic students to focus in different places from their usual workspace. 

A battery of dyslexic students will be selected to test the VR app according to the design decisions taken. These students will receive a VRAIlexia box composed by a VR Cardboard and a controller to be able to give feedback about the VR experience from their mobile devices.

The main objective of the mobile application is to create several VR scenarios (noisy class, natural landscape, diaphanous room, and infinite room) by which dyslexic students can carry out psychometric tests, based on silent reading performance, combined with the Rosenberg scale to record self-esteem of users.

Seven prototypes have been built in the last year (April 2021-May 2022) for the Out of the Box app with the following main features:

\begin{enumerate}
    \item Environment test: visual test for the different environments defined: class, nature, and different size rooms.
    \item Controller test: interaction with the controller and the different VR objects and tracking of the point of view.
    \item Usability test: evaluation of the pre-test interactions (tutorial), user interface according to the number of actions and feedback, and in-game support.
    \item Rosenberg test: representation of the emotional states and legibility of the questions according to the environment.
    \item Silent reading test: button panel representation, voice recognition, and interaction with the text and during the reading.
    \item Database test: online database test for both Rosenberg and Silent Reading.
    \item Integration test: pre-test tutorial, project introduction, silent reading, and Rosenberg test modules are combined and tested as a group.
\end{enumerate}

The app database consists of six different tables: users, environments, languages, silent reading results, Rosenberg results and emotional states.
The personal data corresponding to the dyslexic students is stored in table users. The controlled environment where each test is done is described in table environments. The first language of the dyslexic student is presented in table languages. Results and measurements corresponding to the silent reading test \cite{BDA} and to the Rosenberg test are presented in their corresponding tables. Finally, the description of the different states corresponding to dyslexic students for the Rosenberg test can be retrieved from table emotional states.

The list of fields corresponding to personal data stored in the database includes: Id (unique identifier per user); name; surname; age; gender; e-mail; associated learning difficulties (dysorthography, dyscalculia, dysgraphia, others); additional difficulties related to dyslexia that the students suffered and finally the registration date. As previously mentioned, in the database are included other factors such as environment description and languages (English, Italian, Spanish, or French). 

The Out of the Box app includes two different tests behind two virtual doors:

\begin{itemize}
    \item First test: results and measurements corresponding to the silent reading test containing five types of interactions (selecting the button to be pushed according to the color, following a sequence of buttons, keeping the button pushed and releasing when ordered, selecting a word within the text and voice recognition) (Fig. 1). The results are stored in the database as: \begin{enumerate}
        \item Starting time.
        \item Number of errors (from 0 to 9).
        \item Time per interaction (nine different fields).
        \item Environment (4 different environments).
        \item Voice recognition error (microphone errors).
    \end{enumerate}
    
    \begin{figure}
        \centering
        \begin{subfigure}[b]{0.47\textwidth}
             \centering
             \includegraphics{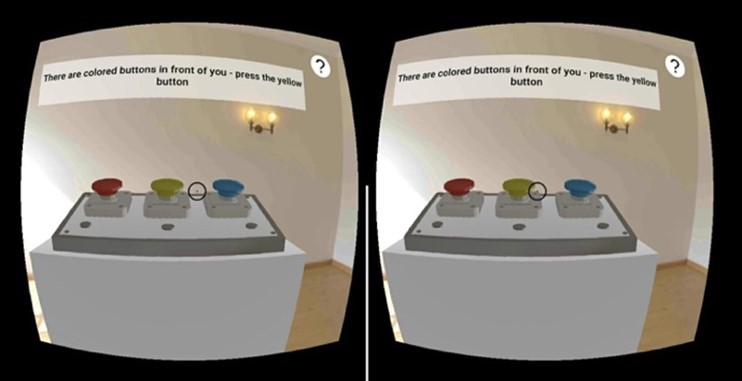}
             \caption{Diaphanous room environment.}
        \end{subfigure}
        \begin{subfigure}[b]{0.47\textwidth}
             \centering
             \includegraphics{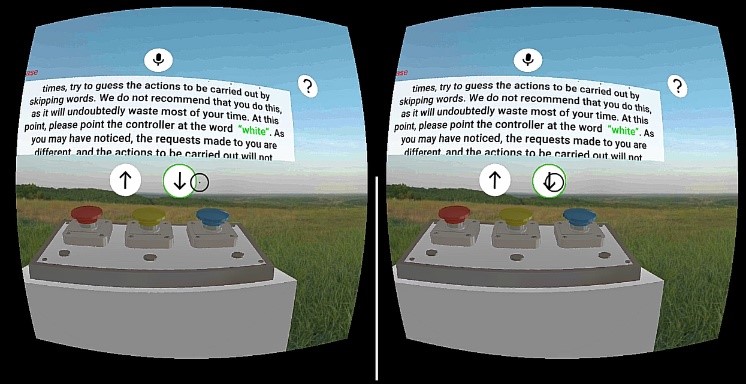}
             \caption{Natural landscape environment.}
        \end{subfigure}

        \caption{Silent test. Instructions immersed in the text are provided to the user in different environments: diaphanous room (a) and natural landscape (b).}
        \label{fig:1}
    \end{figure}
    
    \item Second test: results and measurements corresponding to the Rosenberg test (Fig. 2) containing five types of different states to be selected. The Rosenberg test is a scale of 10 items, whose results depends on the value the of each affirmation from "strongly agree" to "strongly disagree". The score is structure from high levels, medium or normal and low levels of self-esteem \cite{rosenberg}. These selected states are stored in the database as: \begin{enumerate}
        \item Starting time.
        \item Elapsed time.
        \item Environment (4 different environments).
        \item Emotional state selected for each question (ten different questions). The results are based on the description of the different states corresponding to dyslexic students for the Rosenberg test, coding the name of the state (e.g., low level) and description according to cut-off points. The final score of the test is obtained by considering \cite{rosenberg}: 
        \begin{itemize}\item The Rosenberg test is a scale of 10 items, whose results are scored from 30 to 40, codified as a high level of self-esteem; score from 26 to 29, is coded as medium levels; and finally, a score lower than 25 points, is equal to low levels of self-esteem, which implies that this person has severe problems regarding their self-esteem.\end{itemize}

    \end{enumerate}
    
    \begin{figure}[H]
        \centering
        \includegraphics[width=0.47\textwidth]{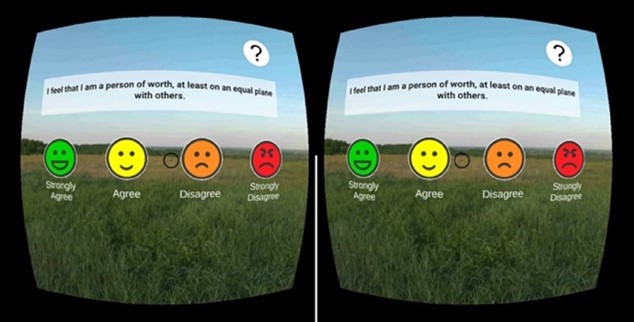}
        \caption{Rosenberg test for determining emotional state.}
        \label{fig:2}
    \end{figure}
    
\end{itemize}

This information from the dataset will be used to extract knowledge by the researchers from the VRAIlexia project to determine the links and associations between the results from both tests and other data, such as individual variables for each user, time taken per action/answer, or the number of correct answers.

\subsection{AI-based assistant for dyslexic students}
The AI module aims to determine which tools and strategies are the most useful for each dyslexic student. A total of seventeen different tools and twenty-two different strategies have been considered to predict their usefulness in reducing the effect of some of the student problems caused by dyslexia.

The AI module's input composes of the information extracted from the previously defined virtual reality application, the surveys (available in Italian, Spanish, and French) carried out by students with dyslexia, and some information obtained from their clinical analyses. This data are going to be put together and preprocessed to obtain a suitable dataset for training our AI models.

\begin{table*}
\centering
\caption{Difficulties, support tools and learning strategies considered.}
\label{tab:info_data}
\begin{tabular}{|c|c|c|c|}
\hline
\textbf{ID} & \textbf{Difficulty/Tool/Strategy}                                        & \textbf{ID} & \textbf{Difficulty/Tool/Strategy}                                       \\ \hline
P1          & Reading                                                    & T15         & Audio recording of lessons                                  \\
P2          & Writing                                                    & T16         & Video lessons                                               \\
P3          & Understanding difficult words                              & T17         & Supplementing study material with internet research         \\
P4          & Understanding the lessons                                  & S1          & A person reading for him/her                                \\
P5          & Concentration                                              & S2          & A map made by himself/herself                               \\
P6          & Paying attention during presential lessons                 & S3          & A scheme made by himself/herself                            \\
P7          & Paying attention during online lessons                     & S4          & A summary made by himself/herself                           \\
P8          & Memorising recently studied concepts                       & S5          & Repeat the studied material                                 \\
P9          & Remembering concepts studied during the exam               & S6          & Marking keywords                                            \\
P10         & Study time management                                      & S7          & Underlining with different colours                          \\
P11         & Taking notes                                               & S8          & Having a study group                                        \\
P12         & Limited time available to prepare a task/question/exam     & S9          & Having a tutor                                              \\
T1          & Human voice audio book                                     & S10         & Dyslexic student group to exchange resources                \\
T2          & Robotic voice audio book                                   & S11         & Presential lessons                                          \\
T3          & Different colour words                                     & S12         & Online lessons available                                    \\
T4          & Using the EasyReading font                                 & S13         & Taking breaks during lessons                                \\
T5          & Using a smart pen or tablet to take notes and record voice & S14         & Lesson slides available                                     \\
T6          & Clearer layout of the study material                       & S15         & Recording the lesson                                        \\
T7          & Having the key words of the text highlighted               & S16         & Taking notes                                                \\
T8          & Prepared concept maps                                      & S17         & Having the lesson plan in advance                           \\
T9          & Prepared schemes                                           & S18         & Dividing an examination/task/question into several parts    \\
T10         & Prepared summaries                                         & S19         & Only written tests                                          \\
T11         & E-Books                                                    & S20         & Only oral tests                                             \\
T12         & Digital tutor                                              & S21         & Conducting the exams in the presence of the professor alone \\
T13         & Images to help understand the meaning of difficult words   & S22         & Having an online database with notes made by other students \\
T14         & Images that help to memorise a concept                     &             &                                                             \\ \hline
\end{tabular}
\end{table*}

The VR application has been already distributed to the different countries, but a complete collection of students’ data has not yet been obtained. Therefore, to demonstrate the performance of the proposed AI tool, we conducted this first experimentation over a dataset previously collected in the Italian universities. This dataset is composed of the responses to 52 questions from 719 dyslexic students. The first 12 questions were about how different problems occasioned by dyslexia (reading comprehension, writing problems, concentration difficulties, etc.) affect to the student. In the following 17 questions, students have to indicate how useful a certain learning support tool (concept maps, words in different colors, etc.) is for them.

Finally, the other 22 questions were related to different learning strategies (someone reading  for them, a scheme done by them, study groups, etc.) and how effective they are for the students when studying. The responses to these questions were collected on a Likert scale from 0 to 5, where 0 means the student has fewer problems or a tool is less useful and 5 the difficulties are greater, or a tool is very useful. The different difficulties, tools, and strategies can be seen together with their identifiers in Table \ref{tab:info_data}. For more details on the survey items, please see \cite{juri}. In addition, after a preliminary preprocessing step, the question related to the tool “Using EasyReading font” (identified as T4) was discarded from the dataset due to most students do not answer it, resulting in 38 features for our final dataset.

\section{Experimental design and analysis of results}

This section describes the different experimental processes conducted using the tools developed during this project. Once the dataset has been described in the previous section, the experimental design is presented in more depth and finally the analysis of the obtained results is discussed.

During the experiments, all the different tools and strategies were considered as different output variables. The objective is to obtain the best model to predict each of these output attributes. A first experimentation considering the dataset as a multilabel classification problem was contemplated. However, due to a lack of interest in the study of the relationships between the output variables, we decided to divide the problem into 38 different single-class problems considering each output individually. All these problems will have the same twelve input variables, representing the difficulties of the dyslexic students, and only one variable to predict, which could be a support tool or a learning strategy. Moreover, a “self-esteem” threshold will be considered to predict if a tool or strategy is useful for dyslexic students. The output values in the Likert scale will be considered as 1 (useful tool or strategy) if they are greater than this threshold and as 0 (not useful tool or strategy) otherwise. In addition, in some of the experiments, this threshold has also been applied to binarise the inputs, to check if easier inputs help the training of some of the different models. Finally, we have designed a consensus submodule for trying to improve the classification of some specific tools and strategies. This consensus decides the final predicted label of each pattern by the most frequently predicted one among the best models of some of the algorithms.

We have considered four different algorithms to construct the models to predict the objective class: Random Forest (RF) \cite{9}, K-Nearest Neighbors (KNN) \cite{10}, Support Vector Machines (SVM) \cite{11}, and Logistic Regression \cite{12}. The documentation of these algorithms can be found in \cite{13}. The final objective of the experimentation is to obtain the best model and the configuration to predict each of the tools and strategies. The different obtained models will be compared for each output according to the Correct Classification Rate metric (\ref{eq:ccr}). Moreover, for a fair comparison, the experimentation results for each output, algorithm, and configuration will be obtained using 10-fold cross-validation.

\begin{equation}
    CCR = \frac{Correctly\:classified\:patterns}{Total\:number\:of\:patterns}
    \label{eq:ccr}
\end{equation}

Next, we present and discuss some of the most interesting results obtained by the different experiments conducted during this work. Specifically, we will focus on the results obtained by AI in predicting useful tools and strategies for dyslexic students.

Table \ref{tab:tools} and Table \ref{tab:strategies} show the highest scores obtained by the best-achieved model for each tool and strategy, respectively. We have specified the applied configuration for reaching this accuracy, including the self-esteem threshold (Thr), the input type, and if the consensus (Cons) was used to improve the final classification.

\begin{table}
\centering
\caption{Best model to predict each support tool.}
\label{tab:tools}
\begin{tabular}{|c|c|c|c|c|c|}
\hline
\textbf{ID} & \textbf{Best Model}        & \textbf{CCR}  \\ \hline
T1            & SVM RBF                  & 0.7443 \\ \hline
T2            & RF, 50 estimators        & 0.9433 \\ \hline
T3            & SVM Linear                & 0.9111 \\ \hline
T5            & SVM RBF                  & 0.8852 \\ \hline
T6            & KNN K=7                   & 0.9538 \\ \hline
T7            & SVM Linear                & 0.9761 \\ \hline
T8            & KNN K=11                  & 0.9325 \\ \hline
T9            & SVM RBF                   & 0.9298 \\ \hline
T10           & SVM RBF                  & 0.9436 \\ \hline
T11           & SVM RBF                  & 0.7246 \\ \hline
T12           & SVM RBF                  & 0.7410 \\ \hline
T13           & KNN K=9                  & 0.9449 \\ \hline
T14           & SVM RBF                & 0.9633 \\ \hline
T15           & SVM Linear                & 0.9354 \\ \hline
T16           & RF 50 estimators          & 0.9279 \\ \hline
T17           & SVM linear                & 0.9367 \\ \hline
\end{tabular}
\end{table}

\begin{table}
\centering
\caption{Best model to predict each learning strategy.}
\label{tab:strategies}
\begin{tabular}{|c|c|c|c|c|c|}
\hline
\textbf{ID} & \textbf{Best Model}        & \textbf{CCR}  \\ \hline

S1            & LR                        & 0.7764 \\ \hline
S2            & RF 50 estimator           & 0.9689 \\ \hline
S3            & SVM Linear                & 0.979  \\ \hline
S4            & KNN K=5                   & 0.9666 \\ \hline
S5            & RF, 50 estimators         & 0.9836 \\ \hline
S6            & RF, 50 estimators         & 0.9738 \\ \hline
S7            & LR                       & 0.9403 \\ \hline
S8            & SVM RBF                   & 0.8902 \\ \hline
S9            & LR                        & 0.8787 \\ \hline
S10           & RF, 50 estimators        & 0.9016 \\ \hline
S11           & RF, 50 estimators        & 0.9443 \\ \hline
S12           & LR                        & 0.9636 \\ \hline
S13           & SVM Linear                & 0.9846 \\ \hline
S14           & SVM Linear                & 0.9898 \\ \hline
S15           & SVM Linear               & 0.9603 \\ \hline
S16           & RF, 50 estimators         & 0.9725 \\ \hline
S17           & RF, 50 estimators         & 0.9551 \\ \hline
S18           & RF, 50 estimators         & 0.9698 \\ \hline
S19           & RF, 50 estimators        & 0.8164 \\ \hline
S20           & RF, 50 estimators         & 0.8554 \\ \hline
S21           & RF, 50 estimators         & 0.8626 \\ \hline
S22           & RF, 50 estimators         & 0.9390 \\ \hline
\end{tabular}
\end{table}
\begin{table}
\centering
\caption{Best model to predict each support tool.}
\label{tab:tools}
\begin{tabular}{|c|c|c|c|c|c|}
\hline
\textbf{ID} & \textbf{Best Model}        & \textbf{Thr} & \textbf{Input}   & \textbf{Cons} & \textbf{Score}  \\ \hline
T1            & SVM RBF           & 4               & Numeric & Yes       & 0.7443 \\ \hline
T2            & RF, 50 estimators & 4               & Numeric & No        & 0.9433 \\ \hline
T3            & SVM Linear        & 1               & Binary  & No        & 0.9111 \\ \hline
T5            & SVM RBF           & 1               & Binary  & Yes       & 0.8852 \\ \hline
T6            & KNN K=7           & 1               & Binary  & No        & 0.9538 \\ \hline
T7            & SVM Linear        & 1               & Binary  & No        & 0.9761 \\ \hline
T8            & KNN K=11          & 1               & Numeric & No        & 0.9325 \\ \hline
T9            & SVM RBF           & 1               & Binary  & No        & 0.9298 \\ \hline
T10           & SVM RBF           & 1               & Binary  & No        & 0.9436 \\ \hline
T11           & SVM RBF           & 4               & Binary  & Yes       & 0.7246 \\ \hline
T12           & SVM RBF           & 4               & Binary  & Yes       & 0.7410 \\ \hline
T13           & KNN K=9           & 1               & Binary  & No        & 0.9449 \\ \hline
T14           & SVM RBF           & 1               & Binary  & Yes       & 0.9633 \\ \hline
T15           & SVM Linear        & 1               & Binary  & No        & 0.9354 \\ \hline
T16           & RF 50 estimators  & 1               & Numeric & No        & 0.9279 \\ \hline
T17           & SVM linear        & 1               & Binary  & No        & 0.9367 \\ \hline
\end{tabular}
\end{table}

\begin{table}
\centering
\caption{Best model to predict each learning strategy.}
\label{tab:strategies}
\begin{tabular}{|c|c|c|c|c|c|}
\hline
\textbf{ID} & \textbf{Best Model}        & \textbf{Thr} & \textbf{Input}   & \textbf{Cons} & \textbf{Score}  \\ \hline

S1            & LR                & 4               & Numeric & No        & 0.7764 \\ \hline
S2            & RF 50 estimator   & 1               & Numeric & No        & 0.9689 \\ \hline
S3            & SVM Linear        & 1               & Binary  & No        & 0.979  \\ \hline
S4            & KNN K=5           & 1               & Numeric & No        & 0.9666 \\ \hline
S5            & RF, 50 estimators & 1               & Numeric & No        & 0.9836 \\ \hline
S6            & RF, 50 estimators & 1               & Numeric & No        & 0.9738 \\ \hline
S7            & LR                & 1               & Numeric & No        & 0.9403 \\ \hline
S8            & SVM RBF           & 1               & Binary  & No        & 0.8902 \\ \hline
S9            & LR                & 1               & Numeric & No        & 0.8787 \\ \hline
S10           & RF, 50 estimators & 1               & Numeric & Yes       & 0.9016 \\ \hline
S11           & RF, 50 estimators & 1               & Numeric & No        & 0.9443 \\ \hline
S12           & LR                & 1               & Numeric & No        & 0.9636 \\ \hline
S13           & SVM Linear        & 1               & Binary  & No        & 0.9846 \\ \hline
S14           & SVM Linear        & 1               & Binary  & No        & 0.9898 \\ \hline
S15           & SVM Linear        & 1               & Binary  & No        & 0.9603 \\ \hline
S16           & RF, 50 estimators & 1               & Numeric & No        & 0.9725 \\ \hline
S17           & RF, 50 estimators & 1               & Numeric & No        & 0.9551 \\ \hline
S18           & RF, 50 estimators & 1               & Numeric & No        & 0.9698 \\ \hline
S19           & RF, 50 estimators & 1               & Numeric & Yes       & 0.8164 \\ \hline
S20           & RF, 50 estimators & 1               & Numeric & No        & 0.8554 \\ \hline
S21           & RF, 50 estimators & 1               & Numeric & No        & 0.8626 \\ \hline
S22           & RF, 50 estimators & 1               & Numeric & No        & 0.9390 \\ \hline
\end{tabular}
\end{table}

Looking at the results, we can see that very different configurations have obtained the best results for the different tools and strategies. However, we can also check that some of the algorithms work better with a specific type of input and that they have reached the highest score in very different tools and strategies. These are the cases of the SVM using a RBF kernel, specially for predicting the tools, or the case of the RF using 50 estimators, specially for predicting the strategies. Moreover, we can see that some of the considered configurations have never obtained a good result, so probably they could be discarded in future experiments.

On the other hand, we can see that the highest scores have been reached using self-esteem thresholds of 1 and 4. This causes the problem to have some class imbalance, which could potentially harm the performance of the trained models. This problem will be considered to avoid during the next project stage with a more complete dataset.

Figure \ref{fig:tools} and Figure \ref{fig:strategies} show bar diagrams comparing the achieved CCR scores for the different support tools and learning strategies. Putting it all together, our different models have obtained an 89.96\% mean accuracy for predicting the support tools, reaching more than 90\% in 12  of the 16 considered tools; and a 93.06\% mean accuracy for predicting the learning strategies, reaching more than 90\% in 16 of the 22 considered strategies.

\begin{figure}
    \centering
    \includegraphics[width=0.48\textwidth]{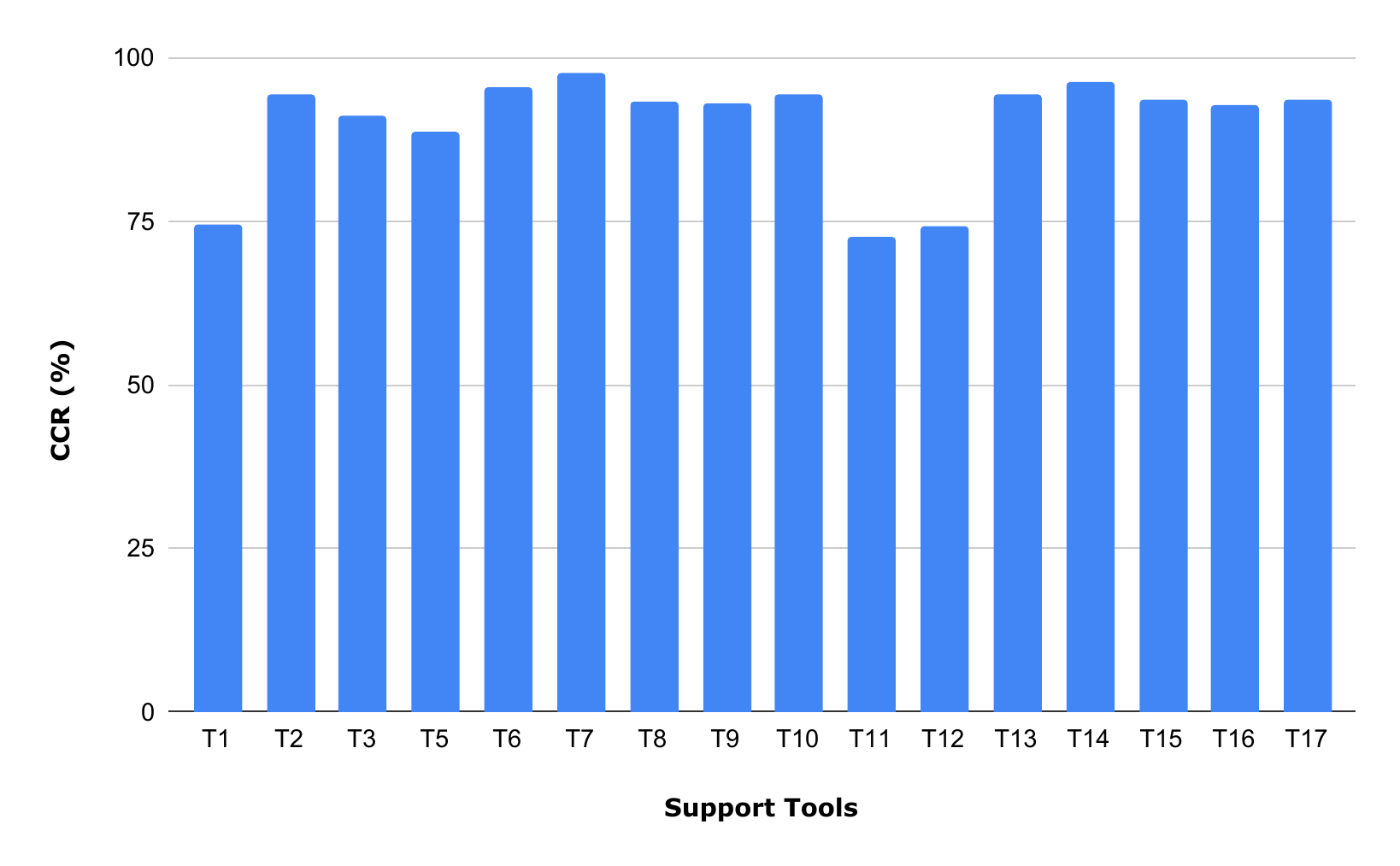}
    \caption{Achieved CCR for predicting the needed supporting tools.}
    \label{fig:tools}
\end{figure}

\begin{figure}
    \centering
    \includegraphics[width=0.48\textwidth]{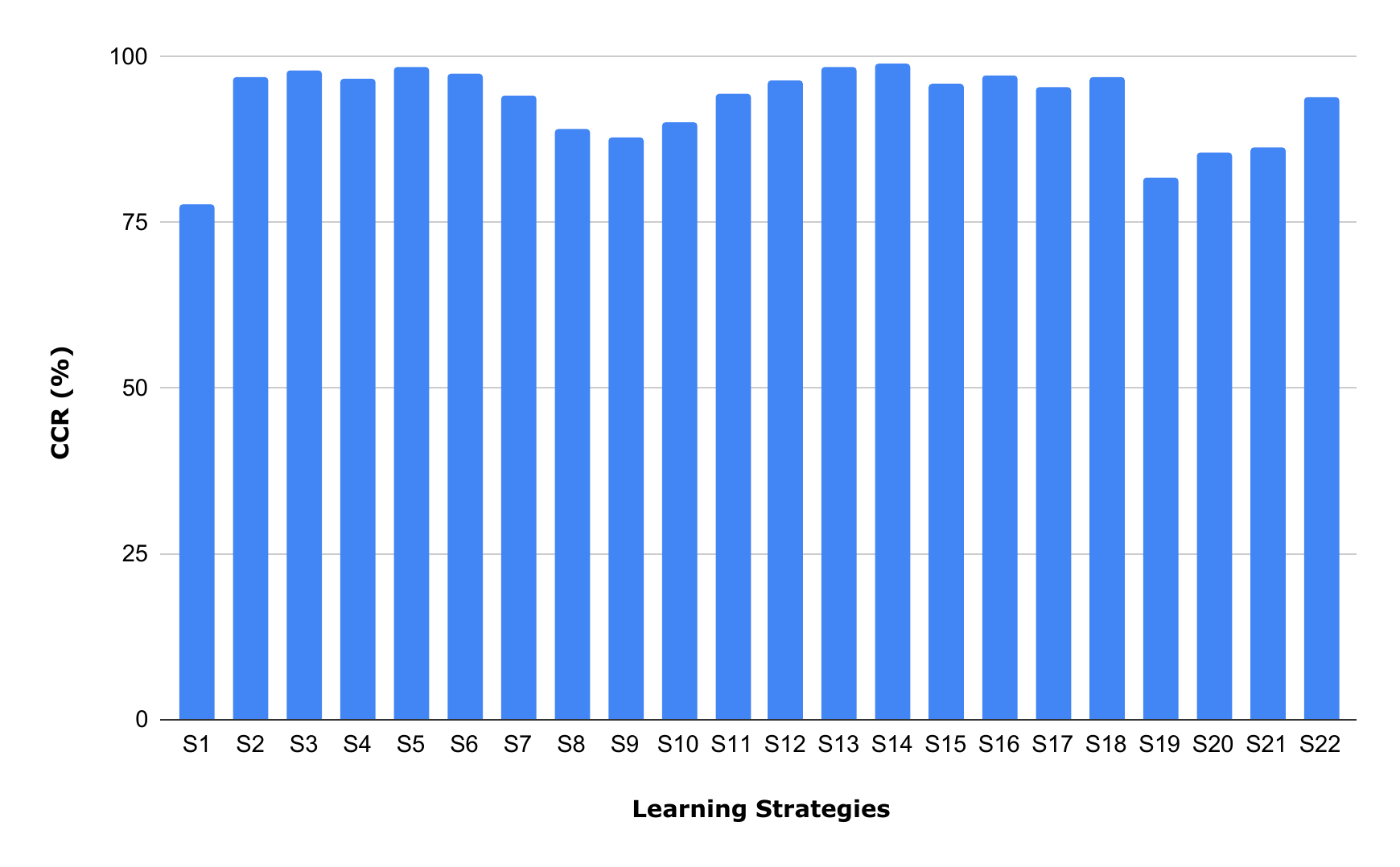}
    \caption{Achieved CCR for predicting the needed learning strategies.}
    \label{fig:strategies}
\end{figure}


\section{Conclusions and future work}

Since students with dyslexia encounter several difficulties during their university career, with a higher risk of dropout respect to other students, tools for students with dyslexia are highly relevant. Nonetheless, the current technological tools mainly focused on primary and secondary school children, leaving students in HEIs supportless. In the higher educational level, the VRAIlexia joins the use of virtual reality and artificial intelligence to find the best supporting tools and learning strategies for each dyslexic student.

The VR helps dyslexic students to experience a variety of sensory stimulation and feedback. They can experience real places, as they see and hear true sights and sounds of a particular environment for the psychometric tests: Silent Reading and Rosenberg tests.

The AI results show how the models have been obtained to determine which tools and strategies are the most useful for each dyslexic student. A total of seventeen different tools and twenty-two different strategies have been considered to predict their usefulness in reducing the effect of some of the student problems caused by dyslexia. More than 90\% mean accuracy has been obtained in the prediction.  

Further work is needed to collect the data corresponding to the psychometric tests and to apply our AI methodology to the new dataset.

\section{Acknowledgements}

We would like to thank all the partners that participate in this project. These results are framed in “VRAIlexia – Partnering Outside the Box: Digital and Artificial Intelligence Integrated Tools to Support Higher Education Students with Dyslexia” funded by the Erasmus+ Programme2014-2020 – Key Action 2: Strategic Partnership Projects. AGREEMENT n. 2020-1-IT02-KA203-080006.This article has been funded with support from the European Commission.

\bibliographystyle{ieeetr}
\bibliography{bibliography}\clearpage

\end{document}